\documentclass[a4paper,11pt]{article}
\usepackage{pos}
\usepackage{newtxtext,newtxmath}
\usepackage{slashed}
\title{Electromagnetic neutrino properties: new constraints and new effects}
\author*{Alexander Studenikin}

\affiliation[a]{Department of Theoretical Physics, Moscow State University, \\ 119991 Moscow, Russia}

\affiliation[b]{Joint Institute for Nuclear Research,\\
141980 Dubna, Moscow Region, Russia}

\emailAdd{studenik@srd.sinp.msu.ru}

\abstract{The electromagnetic properties of neutrinos have attracted considerable attention from researchers for many decades (see \cite{Giunti:2014ixa} for a review). However, until recently, there was no indication in favour of nonzero electromagnetic properties of neutrinos either from laboratory experiments with ground-based neutrino sources or from observations of astrophysical neutrino fluxes.
The situation changed after the XENON collaboration reported \cite{Aprile:2020tmw} results of the search for new physics with low-energy electronic recoil data recorded with the XENON1T detector. The results show an excess of events
over the known backgrounds in the recoil energy which, as one of the possible explanations, admit the presence of
a sizable neutrino magnetic moment, the value of which is of the order of the existing laboratory limitations. In these short notes we give a brief introduction to neutrino electromagnetic properties and focus on the most important constraints on neutrino magnetic moments, charge radii and millicharges from the terrestrial experiments and astrophysical considerations. The promising new possibilities for constraining neutrino electromagnetic properties in future experiments are also discussed.}

\FullConference{%
  40th International Conference on High Energy physics - ICHEP2020\\
  July 28 - August 6, 2020\\
  Prague, Czech Republic (virtual meeting)
}

\begin{document}
\maketitle

%\section{Introduction}

{\bf Introduction.}
The most general form of the neutrino electromagnetic vertex function \cite{Giunti:2014ixa} is given by
%\begin{equation}
$\Lambda_{\mu}^{ij}(q) =  \left( \gamma_{\mu} - {q}_{\mu}
\slashed{q}/q^{2} \right) \left[ f_{Q}^{ij}(q^{2}) + f_{A}^{ij}(q^{2})
q^{2} \gamma_{5} \right] \nonumber
 - i \sigma_{\mu\nu} q^{\nu} \left[ f_{M}^{ij}(q^{2}) +
i f_{E}^{ij}(q^{2}) \gamma_{5} \right]$ ,
%\end{equation}
where $\Lambda_{\mu}(q)$ and form factors $f_{Q,A,M,E}(q^2)$ are $3\times 3$ matrices in  the space of massive neutrinos.  In the case of coupling with a real photon ($q^2=0$) form factors provide four sets of neutrino electromagnetic characteristics: 1) the dipole magnetic moments $\mu_{ij}=f_{M}^{ij}(0)$,
2) the dipole electric moments $\epsilon_{ij}=f_{E}^{ij}(0)$, 3) the millicharges $q_{ij}=f_{Q}^{ij}(0)$ and
4) the anapole moments $a_{ij}=f_{A}^{ij}(0)$.

%\section{Neutrino dipole magnetic and electric moments}
{\bf Neutrino dipole magnetic moments.} The most well understood and studied among neutrino electromagnetic characteristics are the neutrino magnetic moments.
In the Standard Model with massless neutrinos magnetic moments  of neutrinos are zero. Therefore, it is believed that the studies of neutrino electromagnetic properties open a window to {\it new physics} \cite{Giunti:2014ixa,Aprile:2020tmw,Studenikin:2008bd,Studenikin:2018vnp}. In a minimal extension of the Standard Model  the diagonal magnetic  moment of a Dirac neutrino is given \cite{Fujikawa:1980yx} by
%\begin{equation}
$\mu^{D}_{ii}
  = \frac{3e G_F m_{i}}{8\sqrt {2} \pi ^2}\approx 3.2\times 10^{-19}
  \Big(\frac{m_i}{1 \ \mathrm{eV} }\Big) \mu_{B}$,
%  \end{equation}
$\mu_B$ is the Bohr magneton. The Majorana neutrinos can have  only transition
(off-diagonal) magnetic
moments  $\mu^{M}_{i\neq j}$. The same is valid also for the flavour neutrinos in the case of the Majorana mass states.

The most stringent constraints on the neutrino
magnetic moments are obtained with the reactor antineutrinos
(GEMMA Collaboration \cite{GEMMA:2012}):
%\begin{equation}\label{GEMMA}
$\mu_{\nu_e} < 2.9 \times 10^{-11} \mu_{B}$,
%\end{equation}
and solar neutrinos (Borexino Collaboration \cite{Borexino:2017fbd}):
%\begin{equation}\label{Borexino}
${\mu}_\nu ^{eff}< 2.8 \times
10^{-11} \mu _B$. The last limit can be translated to the upper limits for flavour neutrinos: $(\mu_{\nu_e}, \mu_{\nu_{\mu, \tau}}) \sim (4, 6 ) \times 10^{-11} \mu _B$.
%\end{equation}

Note that in general in the scattering experiments the neutrino
is created at some distance from the detector as a flavor neutrino, which is a
superposition of massive neutrinos. Therefore, the magnetic
and electric moments that are measured in these experiments are not that of a
massive neutrino, but there are effective moments that take into account the neutrino mixing and oscillations
during the propagation between the
source and detector \cite{Grimus:1997aa, Beacom:1999wx}.
For the recent and detailed study of the neutrino electromagnetic characteristics
dependence on neutrino mixing see \cite{Kouzakov:2017hbc}.

A new phase of the GEMMA project for measuring the neutrino magnetic moment is now underway at the Kalinin Power Plant in Russia. The discussed next experiment \cite{Belov:2015ufh} called GEMMA-3$ / \nu  $GEN  is aimed at the further increase in sensitivity to the neutrino magnetic moment and will reach the level of
%\begin{equation}\label{nuGEN}
$\mu_{\nu_e} \sim (5{-}9) \times 10^{-12}\mu _B $.
%\end{equation}
To reach the  claimed limit on the neutrino magnetic moment  the $\nu$GEN experiment setup reasonably improves characteristics in respect to those of the previous editions of the GEMMA project. The most important are the following \cite{Lubashevskiy:2020}: 1) a factor of 2 increase in the total neutrino flux at the detector because of much closer location of the detector to the reactor core, 2) a factor of 3.7 increase in the total mass of the detector, 3) the energy threshold would be improved from $2.8 \ keV$  to $ 200 \ eV$. Furthermore, the $\nu$GEN experimental setup is located in the new room at the Kalinin Power Plat with much better (by an order of magnitude) gamma-background conditions and on a moveable platform. The later gives an opportunity to vary online the neutrino flux and thus suppress systematic errors.

The observation of coherent elastic neutrino-nucleus scattering reported for the first time \cite{Akimov:2017ade} by the
COHERENT experiment at the Spallation Neutron Source can be also used for constraining neutrino electromagnetic properties. For the case of neutrino magnetic moments, however, as it was shown in \cite{Kosmas:2017tsq}
and then confirmed in recent studies (see, for instance, \cite{Miranda:2020tif}  ) the bounds for the flavour neutrino magnetic moments are of the order $\mu_e , \mu_\mu \sim 10^{-8} \mu _{B}$.

In the recent studies \cite{Miranda:2020kwy} it is shown that the puzzling results of the XENON1T collaboration \cite{Aprile:2020tmw} at few keV electronic recoils  could be due to the scattering of solar neutrinos endowed with finite Majorana transition magnetic moments of the strengths lie within the limits set by the Borexino experiment  with solar neutrinos \cite{Borexino:2017fbd}. The comprehensive analysis of the existing and new extended mechanisms for enhancing neutrino transition magnetic moments to the level appropriate for the interpretation of the XENON1T data  and leaving neutrino masses within acceptable values is provided in \cite{Babu:2020ivd}.

In the most recent paper \cite{Cadeddu:2019qmv} we have proposed an experimental setup to observe coherent elastic neutrino-atom scattering using electron antineutrinos from tritium decay and a liquid helium target. In this scattering process with the whole atom, that has not beeen observed so far, the electrons tend to screen the weak charge of the nucleus as seen by the electron antineutrino probe.
 Finally, we study the sensitivity of this apparatus to a possible electron
 neutrino magnetic moment and we find that it is possible
 to set an upper limit of about
% \begin{equation}
$\mu_{\nu} < 7 \times 10^{-13} \mu_{B}$,
%\end{equation}
%at 90 \%  C.L.,
that is more than one order of magnitude smaller than
the current experimental limits from GEMMA and Borexino.

An astrophysical bound on an effective neutrino magnetic moment (valid for both cases of
Dirac and Majorana neutrinos) is provided
\cite{Raffelt-Clusters:90, Viaux-clusterM5:2013, Arceo-Diaz-clust-omega:2015}
by observations of the properties of globular cluster stars:
%\begin{equation}
$\Big( \sum _{i,j}\left| \mu_{ij}\right| ^2\Big) ^{1/2}\leq (2.2{-}2.6) \times
10^{-12} \mu _B$. There is also a statement \cite{deGouvea:2012hg}, that
observations of supernova fluxes  in the future largevoluem experiments like JUNO, DUNE and Hyper-Kamiokande ( see for instance
\cite{An:2015jdp,Giunti:2015gga,Lu:2016ipr}) may reveal the effect of  collective  spin-flavour oscillations  due to the Majorana neutrino transition moment $\mu^{M}_\nu \sim 10^{-21}\mu_B$. Other new possibilities for neutrino
magnetic moment visualization in extreme astrophysical environments are
considered recently in \cite{Grigoriev:2017wff,Kurashvili:2017zab}.
%\end{equation}

A general and termed model-independent upper bound on the Dirac neutrino
magnetic moment, that can be generated by an effective theory beyond
a minimal extension of the Standard Model, has been derived in
\cite{Bell:2005kz}: $\mu_{\nu}\leq
10^{-14}\mu_B$. The corresponding limit for transition moments of Majorana neutrinos is much weaker \cite{Bell:2006wi}.

{\bf Neutrino dipole electric moments.} In the theoretical framework with $CP$ violation a neutrino
can have nonzero electric moments $\epsilon_{ij}$. In the laboratory neutrino
scattering experiments for searching $\mu_{\nu}$ (for instance, in the GEMMA experiment)
the electric moment $\epsilon_{ij}$ contributions interfere with
those due to $\mu_{ij}$. Thus, these kind of experiments also provide constraints
on $\epsilon_{ij}$. The astrophysical bounds on $\mu_{ij}$
are also applicable for constraining $\epsilon_{ij}$ (see \cite{Raffelt-Clusters:90, Viaux-clusterM5:2013, Arceo-Diaz-clust-omega:2015} and \cite{Raffelt:2000kp}).

%\section{Neutrino electric millicharge}
{\bf Neutrino electric millicharge.} There are extensions of the Standard Model that allow for nonzero
neutrino electric millicharges. This option can be provided by
not excluded experimentally possibilities for hypercharhge dequantization or
another {\it new physics} related with an additional $U(1)$ symmetry
peculiar for extended theoretical frameworks. Note that neutrino millicharges
are strongly constrained on the level $q_{\nu}\sim 10^{-21} e_0$
($e_0$ is the value of an electron charge) from neutrality of the hydrogen atom.

 A nonzero neutrino millicharge $q_{\nu}$ would contribute to the neutrino electron scattering in the terrestrial experiments. Therefore, it is possible to get bounds on $q_{\nu}$ in the reactor antineutrino
 experiments. The most stringent reactor antineutrino constraint
 %\begin{equation}
 $q_{\nu}< 1.5 \times 10^{-12} e_0 $
 %\end{equation}
 is obtained in \cite{Studenikin:2013my} within the free-electron approximation using the GEMMA experimental data \cite{GEMMA:2012}. This limit is cited by the Particle Data Group since 2016 (see also \cite{Zyla:2020zbs}).
A certain increase in the cross section is expected in the case when instead of the free-electron approximation one accounts for
the so called atomic ionization effect \cite{Chen:2014dsa}, and the obtained corresponding limit on the neutrino millicharge is
%\begin{equation}
 $q_{\nu} < 1 \times 10^{-12} e_0$.
 %\end{equation}

 The expected increasing sensitivity to the neutrino-electron scattering of the future $\nu$GEN experiment that is aimed to reach a new  limit  for the magnetic moment would provide a possibility \cite{Studenikin:2013my} to check the neutrino millicharge at the scale of  $ q_{\nu} \sim  10^{-13} e_0$.

As it has been already mentioned above, the coherent elastic neutrino-nucleus scattering \cite{Akimov:2017ade} is a new
powerful tool to probe the electromagnetic neutrino properties \cite{Kosmas:2017tsq}. In the flavour basis neutrinos can have diagonal $q_{lf}$ ($l=f$, $l,f = e, \mu, \tau$) and transition $q_{lf}$ ($l\neq f $) electric charges (see, for instance, \cite{Giunti:2014ixa} and \cite{Kouzakov:2017hbc}). Such possibilities are not excluded by theories beyond the Standard Model. Recently \cite{Cadeddu:2019eta} from the analysis of the COHERENT data new constraints for all neutrino charges on the level of $\sim 10^{-7} e_0$ are obtained. It follows, that the bounds for involving the electron neutrino flavour charges $q_{ee}, q_{e\mu}$ and $q_{e\tau}$ are not competitive with respect to constraints $\sim 10^{-12} e_0$ obtained for the effective electron neutrino charge $q_{eff}= \sqrt{q_{ee}^2 +q_{e\mu}^2 +q_{e\tau}^2}$ from the reactor antineutrino scattering experiments \cite{Studenikin:2013my, Chen:2014dsa}. Note, that the bounds for $q_{\mu \mu}$ and $q_{\mu \tau}$ from a laboratory data are obtained in \cite{Cadeddu:2019eta} for the first time.

The most recent and one of the most detailed statistical studies \cite{Parada:2019gvy} of experimental data from the elastic neutrino-electron and coherent neutrino-nucleus scattering show that the combined inclusion of different experimental data can lead to stronger
constraints on $q_\nu$ than those based on individual analysis of different experiments.

A neutrino millicharge would have specific phenomenological consequences
in astrophysics because of new electromagnetic processes are opened
due to a nonzero charge (see \cite{Giunti:2014ixa,Raffelt:1996wa,Studenikin:2012vi}). Following this line, the most stringent astrophysical constraint on neutrino millicharges
%\begin{equation}\label{q_astr}
$q_{\nu}< 1.3 \times 10^{-19} e_0 $
%\end{equation}
 was obtained in \cite{Studenikin:2012vi}. This bound
follows from the impact of the {\it neutrino star turning} mechanism ($\nu ST$) \cite{Studenikin:2012vi} that can be considered as a {\it new physics} phenomenon end up with a pulsar rotation frequency
shift engendered by the motion of escaping from the
star neutrinos along curved trajectories due to millicharge interaction with a constant
magnetic field of the star. The existed other astrophysical constraints on the neutrino millicharge, however less restrictive than that of \cite{Studenikin:2012vi}, are discussed in \cite{Giunti:2014ixa,Parada:2019gvy}.

%\section{Neutrino charge radius and anapole moment}
{\bf Neutrino cherge radius and anapole moment.} Even if a neutrino millicharge is vanishing, the electric form factor
$f^{ij}_{Q}(q^{2})$ can still contain nontrivial information about
neutrino electromagnetic properties. The corresponding electromagnetic characteristics is
determined by the derivative of $f^{ij}_{Q}(q^{2})$ over $q^{2}$  at
$q^{2}=0$ and is termed neutrino charge radius,
%\begin{equation}
%\label{nu_cha_rad_1}
$\langle{r}_{ij}^{2}\rangle
=-
6
%\left.
\frac{df^{ij}_{Q}(q^{2})}{dq^{2}} \
_{\mid _ {q^{2}=0}}
$ (see \cite{Giunti:2014ixa} for the detailed discussions).
%\end{equation}
Note that for a massless neutrino the neutrino charge radius is the only
electromagnetic characteristic that can have nonzero value. In the Standard Model
the neutrino charge radius and the anapole moment are not defined separately,
and there is a relation between these two values: $a = - \frac{\langle{r}^{2}\rangle}{6}$.

A neutrino charge radius contributes to the neutrino scattering cross section on electrons and thus
can be constrained by the corresponding laboratory experiments \cite{Bernabeu:2004jr}.
In all but one previous studies it was claimed
 that the effect of the neutrino
charge radius can be included just as a shift of the vector coupling constant $g_V$
in the weak
contribution to the cross section.
However, as it has been recently demonstrated in \cite{Kouzakov:2017hbc} within the direct calculations of
the elastic neutrino-electron scattering cross section accounting for all possible neutrino electromagnetic characteristics
and neutrino mixing, this is not the fact. The neutrino charge radius dependence of the cross section
is more complicated and there are, in particular, the dependence on the interference terms of the type
$g_{V}\langle{r}_{ij}^{2}\rangle$ and also on the neutrino mixing.
The current constraints on the flavour neutrino charge radius $\langle{r}_{e,\mu,\tau}^{2}\rangle\leq 10^{-32} - 10^{-31} \ cm ^2$
from the scattering experiments differ only by 1 to 2
orders of magnitude from the values $\langle{r}_{e,\mu,\tau}^{2}\rangle\leq 10^{-33} \ cm ^2$ calculated within the minimally extended Standard Model with right-handed neutrinos
\cite{Bernabeu:2004jr}. This indicates that the minimally extended Standard Model neutrino charge radii could be experimentally tested in the near future.

Note that there is a need to re-estimate experimental constraints on
$\langle{r}_{e,\mu,\tau}^{2}\rangle$  from the scattering experiments following
new derivation of the cross section \cite{Kouzakov:2017hbc} that properly accounts for the interference of the weak and charge radius electromagnetic interactions and also for the neutrino mixing.

Recently constraints on  charged radii  have been obtained
\cite{Caddedu:2018prd} from the analysis of the data on coherent
elastic neutrino-nucleus scattering obtained in the COHERENT experiment
\cite{Akimov:2017ade,Akimov:2018vzs}. In addition to the customary diagonal
charge radii $\langle{r}_{e,\mu,\tau}^{2}\rangle$
also the neutrino transition (off-diagonal) charge radii have been constrained
in \cite{Caddedu:2018prd} for the first time:
$\left(|\langle r_{\nu_{e\mu}}^2\rangle|,|\langle r_{\nu_{e\tau}}^2\rangle|,|\langle r_{\nu_{\mu\tau}}^2\rangle|\right)
< (22,38,27)\times10^{-32}~{\rm cm}^2$. Since 2018 these limits are included by the Particle Data Group to Review of Particle Properties (see also \cite{Zyla:2020zbs}) and also were noted by the Editors' Suggestion as the most important results (PRD Highlights 2018) published in the journal.
%For the future progress in studying (or constraining) neutrino electromagnetic properties
%a rather promising claim was made in  \cite{deGouvea:2012hg,deGouvea:2013zp}. It was shown that
%even tine values of the Majorana neutrino transition moments
%would probably be tested in future high-precision experiments with the astrophysical neutrinos.

The work is supported by the Russian Foundation for Basic Research under grant No. 20-52-53022-GFEN-a.

\end{document}